\documentclass[12pt]{article}

\input epsf
\def\beq{\begin{eqnarray}}
\def\eeq{\end{eqnarray}}

\def\mpl{M_{\rm Pl}}

\def\L*{{\cal L}_*}
\def\lsim{\mathrel{\rlap{\lower3pt\hbox{\hskip0pt$\sim$}}
     \raise1pt\hbox{$<$}}}         
\def\gsim{\mathrel{\rlap{\lower4pt\hbox{\hskip1pt$\sim$}}
     \raise1pt\hbox{$>$}}}         

\usepackage{amsmath}
\usepackage{amsfonts}

\begin{document}

\begin{titlepage}

\thispagestyle{empty}

\begin{flushright}
{NYU-TH-07/03/01}
\end{flushright}
\vskip 0.9cm

\centerline{\Large \bf Charged Condensation}

\vskip 0.7cm
\centerline{\large Gregory Gabadadze and  Rachel A. Rosen}
\vskip 0.3cm
\centerline{\em Center for Cosmology and Particle Physics}
\centerline{\em Department of Physics, New York University, New York, 
NY, 10003, USA}

\vskip 1.9cm

\begin{abstract}

We consider Bose-Einstein condensation of massive 
electrically charged scalars in a uniform background of charged fermions. 
We focus on the case when the scalar condensate screens the background 
charge, while the  net charge of the system resides on its boundary 
surface. A distinctive  signature of this substance is that the 
photon  acquires a 
Lorentz-violating mass in the bulk of the condensate. Due to this mass, 
the transverse and longitudinal gauge modes propagate with different 
group velocities. We give qualitative arguments 
that at high enough densities and low temperatures a charged system 
of electrons and helium-4 nuclei, 
if held together by laboratory devices or by force of gravity, 
can form such a substance. We briefly discuss possible 
manifestations  of the charged condensate in compact 
astrophysical objects.

\end{abstract}

\vspace{3cm}

\end{titlepage}

\newpage

{\it 1. Introduction and summary.} Consider a sphere enclosing  
massive stable  charged spin-1/2 particles with  number density  
${\bar J}_0$, and stable  massive spin-0 particles of an 
equal but opposite charge. At some high  temperature the  
substance in the sphere could form hot plasma.  With the decreasing  
temperature the opposite charges would ordinarily  form neutral atoms 
of half-integer spins. These atoms  would not 
be able to Bose-Einstein condense because of their 
spin-statistics.

We will discuss in this work  a different sequence of events  that  
could  take place in the above system. In particular, we will show 
that under certain conditions, instead of forming  neutral atoms, the charged 
scalars could themselves condense,  neutralizing by this condensate 
the background charge of the fermions. 

Especially interesting we find the case  when the system has a net 
overall charge to begin with. In this case, although the resulting 
substance is charge  neutral in  the interior of the sphere, the 
net charge will reside on its surface.  The substance in the bulk 
has distinctive properties. We will show in Section 2 
that propagation  of a photon 
in this  substance is rather special. Even at zero temperature, 
the photon acquires a Lorentz non-invariant  mass term. 
The transverse and longitudinal components of the  photon have equal masses; 
the mass squares are  proportional to ${\bar J}_0$  and inversely 
proportional to the charged  scalar mass. However, the group velocities 
of the transverse and  longitudinal  modes are different. 
The longitudinal mode  is similar to a plasmon excitation of cold plasma.
The transverse  modes  of the photon propagate as 
massive states. We will refer to this phase as the charged condensate, 
emphasizing that the charged scalars have  undergone  Bose-Einstein 
condensation, while the background fermions merely play 
the role of charge neutralizers in the bulk of the substance, 
and the net charge of the system is residing on the boundary.

The above mechanism is universal: the gauge field could be a photon or 
any other $U(1)$ field, while  the charged scalar 
could be a fundamental field, or a composite state made of
other particles, in the regime where its  compositness does not matter.
This may have applications in particle physics and condensed matter systems. 

As a concrete example we imagine a reservoir, or a trap, in which negatively 
charged electrons and positively charged helium-4 nuclei, with 
a nonzero net charge,  could be  put together at densities  high 
enough for an average inter-particle separation  to be smaller than 
the size of a helium atom. In this case, the helium atoms would not form.
The results of Section 2 cannot immediately  be 
applied to this case,  since electrons are lighter 
than the helium nuclei. However, we will argue in  Section 3
that if temperature of the system  
is low  enough for the helium de Broglie wavelength to be 
greater than both the average inter-particle separation and 
the Compton wavelength of the massive photon, then the charged helium-4 
nuclei would fall into the condensate. 
Photons, in the bulk of this substance, would propagate  with a delay 
caused by the acquired mass. Such a system would also have a net
surface charge. Quantitative features of this example are discussed in 
Section 3. Our estimate for the temperature 
is within  the range of the low temperatures that have already been 
achieved in experiments on  Bose-Einstein condensation 
of atoms, see, e.g., \cite {BEC}.

In the above example the charged condensate containing droplet
was assumed to be held together by a rigid boundary or 
external fields in a laboratory.
In Section 4 we point out that gravity could play the role 
of the stabilizing force, and briefly discuss possible manifestations 
of the charged condensation  in compact astrophysical objects.  

A few comments on the literature. The pion condensation due to 
strong interactions is well known 
\cite {Migdal}. In this work we discuss  condensations due to 
electromagnetic interactions instead (or in more general case, 
due to some $U(1)$ Abelian interactions).  It was shown in Ref. 
\cite {Linde} that the constant charge density strengthens 
spontaneous symmetry breaking  when the symmetry is already broken 
by the usual Higgs-like nonlinear potential for the scalar. 
In our work the scalar  has  a conventional 
positive-sign mass term.  The fact that the conventional-mass scalar  
could condense  in the charged background was first shown  
in \cite {Kapusta}. However, the system considered 
in \cite {Kapusta} is  neutral, and thus, 
is physically different from the one studied in 
this work (see, brief comments after eq. (4.6) 
in \cite {Kapusta}).  An expanded discussions of the topics covered
in the present work, with other possible applications will be 
presented elsewhere \cite {GGRR}.

\vspace{0.1in}  

{\it 2. Basic mechanism.} We consider a simplest model that 
exhibits the main phenomenon. Let us start with a system in an 
infinite volume and  at zero-temperature.  The classical Lagrangian  
contains a gauge field $A_\mu$, a charged scalar field $\phi$  with a 
right-sign mass term $m_H^2>0$,  and fermions $\Psi^+,\Psi$ with 
mass $m_J$
\beq
{\cal{L}} = -\tfrac{1}{4}F_{\mu\nu}^2 + 
\vert D_{\mu} \phi \vert^2 - m_H^2 \phi^{\ast} \phi +
{\bar \Psi}i\gamma^\mu D_\mu \Psi -m_J  {\bar \Psi} \Psi
+ \mu \Psi^+\Psi \,.
\label{lagr0}
\eeq
The chemical potential $\mu$ is introduced for the global 
fermion number carried by ${\Psi}$'s (e.g., lepton, baryon or other 
number). The covariant derivatives in  
(\ref {lagr0}) are defined as  $\partial_\mu +ig_{\phi}A_\mu$ for 
the scalars, and  $\partial_\mu +ig_{\psi}A_\mu $ for the 
fermions. Their respective charges, $g_{\phi}$ 
and  $g_{\psi}$,  are different in general. For 
simplicity we  assume that $g_{\phi}=-g_{\psi}\equiv -g$.

To study the ground state  it is convenient to introduce the 
following notations for the scalar, gauge field and fermions: 
$\phi = \tfrac{1}{\sqrt{2}} \sigma\, e^{i \alpha}$, 
$B_{\mu} \equiv A_{\mu} + \tfrac{1}{g}\partial_{\mu} \alpha $, and 
$\psi = \Psi e^{-i \alpha}$. In terms of the gauge invariant variables 
$\sigma$, $B_\mu$ and $\psi$ the Lagrangian, takes the form
\beq
{\cal{L}}=-\tfrac{1}{4}F_{\mu\nu}^2 + 
\tfrac{1}{2}(\partial_{\mu}\sigma)^2+
\tfrac{1}{2}g^2 B_\mu^2 \sigma^2- \tfrac{1}{2}m_H^2 
\sigma^2 + {\bar \psi}i\gamma^\mu D_\mu \psi -m_J  {\bar \psi} \psi
+ \mu \psi^+\psi \,,
\label{lagr}
\eeq
where now $F_{\mu\nu}$ and $D$  are  a field-strength  and covariant 
derivative for $B_\mu$, respectively.  

Fermions in (\ref {lagr}) obey  the conventional Dirac equation with a 
nonzero chemical potential. This implies a net fermion number  
in the system,  ${\bar J}_0$. Since the  fermions  are also electrically 
charged, they  set  a background  electric charge density.  
Such charged fermions  would repel each other. In our case, 
however, the charge will be  screened by the charged scalar 
condensate.  One way to see this is to assume that such a 
self-consistent solution  exists, and then check  
explicitly that it satisfied equations of motion, as we will
do it below.  We consider distance scales that are  greater than 
an average separation between 
the fermions, so  that  their  spatial distribution could be assumed to 
be uniform.  Then, the background  charge 
density  due to the fermions could be approximated  as 
${\bar J}_\mu = {\bar J}_0 \delta _{\mu 0}$, where ${\bar J}_0$ is 
a constant.  The magnitude of  the 
latter is related to  the value of the chemical potential $\mu$. 
In particular, a self-consistent solution of the equations of motion 
implies that  $\mu-\langle gB_0 \rangle = E_F$, where $E_F$ denotes the 
Fermi energy of the background fermion sea, and  is related to 
${\bar J}_0$ as follows, $E_F = \sqrt{(3 \pi {\bar J}_0/4)^{2/3}+ m_J^2}$.

The rest of the equations of motion derived  from  (\ref {lagr}) are:
\beq
\partial^\mu F_{\mu\nu} + g^2 B_\nu \sigma^2 = g {\bar J}_\nu \,,~~~~~
\square \sigma = g^2 B_\nu^2 \sigma - m_H^2 \sigma \,.
\label{Name}
\eeq
The Bianchi identity for the first equation in (\ref {Name}), 
$\partial^\nu (B_\nu \sigma^2)=0$, can also be obtained by varying the action 
w.r.t. $\alpha$. For a constant charge density,  
${\bar J}_\mu = {\bar J}_0 \delta _{\mu 0}$, the theory 
with the scalar field  (\ref {lagr0}) admits a static solution 
with constant  $B_0$ and $\sigma$: 
\beq
\langle B_0 \rangle =B_{0c} \equiv {m_H\over g}\,,~~~~~~~
\langle \sigma \rangle = \sigma_c \equiv 
\sqrt{\frac{ {\bar J}_0 }{m_H}} \,.
\label{b0}
\eeq
The charge density stored in the condensate, 
$J^{\rm scalar}_0=-i [\phi^* D_0\phi-(D_0 \phi)^* \phi] 
= -g\sigma^2 B_0$, equals to $-{\bar J}_0$, by virtue of (\ref {b0}). 
Hence, the total charge density 
$J_{\rm total} = {\bar J}_0 +J^{\rm scalar}_0=0$, vanishes.
The ground state is charge-neutral in its bulk. On the other hand, 
a nonzero  $ \langle B_0 \rangle$ in  (\ref {b0}) suggests that 
there must be an uncompensated charge on a surface at infinity, 
as it will be the case (see below).

Before we continue with studies of small perturbations about the 
solution (\ref {b0}), we would like to make four essential comments:

(i) The expression for the gauge  field in (\ref {b0}) scales
as $1/g$, and is non-perturbative in its nature. Moreover, it diverges in 
the limit $m_H\to \infty$. This seeming non-decoupling of 
the charged scalar field results from  the fact that we're dealing 
with a constant background  {\it  charge density} in an infinite volume, 
i.e., with an infinite background charge. It is not surprising 
then, that such a background  is capable of affecting a 
charged state of an arbitrary mass. Moreover, when $m_H$ 
exceeds the fermion mass, our averaging procedure over the background 
charges should not be applicable in general.   

(ii) In regard with the above discussions, 
it is instructive to regularize the problem by considering a finite 
volume ball of a radius $R$. A nonzero  $ \langle B_0 \rangle$ in 
(\ref {b0}) suggests that there must be an uncompensated charge on 
the surface of the ball, which tends to the value, $Q=m_H R/g$, 
as $R\to \infty$. Indeed, such a charge $Q$ could give rise to a constant  
$ \langle B_0 \rangle=m_H/g$ in the interior of the ball,
where $ \langle B_0 \rangle = Q/R$,  in analogy with a static 
potential inside a conducting ball with surface charge $Q$. 
This is indeed what happens in the  present case. These and other 
finite volume effects  are discussed in detail in 
Section 3.

(iii) Unlike for the fermions,  we have not introduced chemical potential for 
the scalars. However, nonzero $\langle gB_0 \rangle$ acts as 
dynamically  induced chemical potential for the 
perturbations of the scalar. Its value in the ground state, 
$\langle gB_0 \rangle =m_H$, is consistent with the expectation
that the chemical potential be equal to the mass of the scalar
in Bose-Einstein condensate. 

In general, we could have introduced 
chemical potential for the charged scalar, $\mu_s$.  
The  above described  condensation mechanism would still 
take place with the  result, $ \langle gB_0 \rangle  = m_H + \mu_s$, and 
$\sigma^2_c= {\bar J}_0/m_H$,  instead of (\ref {b0}).  
The charge density in the condensate
in this case would read,  $-(\mu_s - g  B_0) \sigma^2= -{\bar J}_0$,
ensuring charge neutrality of the  substance 
in its bulk, but  in general there 
would be  a nonzero  surface charge, unless 
$\mu_s =-m_H$ and $\langle gB_0 \rangle =0$. 

(iv) So far our discussions have been classical.  Upon quantization 
the charged condensate can be thought of a zero-momentum 
state with a non-zero occupation number of the 
charged scalar field quanta. It is useful to consider 
small temperature $T$ in the system, in which case 
the de Broglie wavelength of the condensed scalars, 
$\lambda_T\sim ({1/ m_HT})^{1/2}$, will exceeds the 
average inter-particle separation $\sim {\bar J}^{-1/3}_0$. 
Thus, it makes sense to think of the charged condensate, as 
of any other Bose-Einstein condensate,  to be a macroscopically 
occupied mode. The specifics of our case is that this 
macroscopic state of electrically charged scalars can exist even 
when the Compton wavelength of the corresponding massive photon is 
greater than the average interparticle separation between the scalars.
In the bulk of the condensate the charge is balanced by the 
background charge density of fermions.  

\vspace{0.1in}

The uniform fermion background sets a preferred Lorentz frame.
We study the spectrum and propagation of  perturbations 
in this background frame. For this we introduce small 
perturbations of gauge and scalar fields, $b_\mu$ and $\tau$, 
as follows:
\beq
B_\mu = B_{0c} \delta_{\mu 0} + b_\mu(x)\,,~~~~\sigma = \sigma_c + \tau(x)\,.
\label{pert}
\eeq
The Lagrangian density for the perturbations reads
\beq
{\cal L}_{2}= -{1\over 4} f_{\mu\nu}^2 + \tfrac{1}{2}(\partial_{\mu} \tau)^2+
\tfrac{1}{2}g^2 \sigma_c^2 b_\mu^2 + 2gm_H\sigma_c \,b_0 \tau +...
\label{2lagr}
\eeq
Here $f_{\mu\nu}$ denotes the field strength for $b_\mu$, and 
we dropped all the fermionic terms as well as the cubic and quartic 
interaction terms  of $b$'s  and $\tau$. 
The  last term in  (\ref {2lagr})  is Lorentz 
violating. Calculations of the spectrum of the theory is non-trivial but 
straightforward. We briefly summarize the results. 
First,  $b_0$ is not a dynamical field, as it has no 
time derivatives in (\ref {2lagr}). Therefore, it 
can be integrated  out through its equation of motion, 
leaving us with the equations for three polarizations of a  
massive vector $b_j,~~j=1,2,3$, and one scalar $\tau$. 
These constitute four physical degrees of freedom of the theory. 
The transverse part of the vector $b_j$ 
obeys the free equation 
\beq
\label{bj}
(\square +g^2\sigma_c^2)b^T_j=0,~~{\rm where}~~~b^T_j \equiv b_j-
{\partial_j\over \Delta}(\partial_k b_k)\,.
\eeq 
Therefore, the two states of the 
gauge field carried by $b^T_j$ have the following mass 
\beq
m^2_g=g^2\sigma_c^2= g^2 {{\bar J}_0\over m_H}\,.
\label{mass}
\eeq
Moreover, the frequency $ \omega$ and the three-momentum vector
${\bf p}$ of these two states obey the conventional dispersion 
relation,  $\omega^2 = {\bf p}^2+m^2_g$. 

The longitudinal mode of the gauge field $b^L_j$, and the scalar 
$\tau $, on the other hand,  give rise to the following 
Lorentz-violating dispersion relations (valid for $m_g\neq 0$)
\beq
\omega^2_{\pm}= {\bf p}^2+2m_H^2 + {1\over 2}m_g^2 \pm
\sqrt{4{\bf p}^2m_H^2+ (2m_H^2- {1\over 2}m_g^2 )^2}\,.
\label{LVdisp}
\eeq
The r.h.s. of (\ref {LVdisp}) is positive.
Both of these modes have masses which can be 
obtained by putting  ${\bf p}=0$ in (\ref {LVdisp}). One of them 
coincides with (\ref {mass}), and the  other one, 
has the mass squared  equal to $m^2_s=4m^2_H$. 
Interestingly, the group velocities   of the transverse and 
longitudinal modes of the massive vector boson are  different.
For $m_H\gg m_g$, and for 
an arbitrary ${\bf p}$, the fastest ones are the 
transverse modes, they're followed by the scalar, 
and the longitudinal mode is the slowest. 

In the limit $m_H\to 0$, (\ref {LVdisp}) 
describes a massive  longitudinal component of a vector 
bosons of mass $m_g$, and a massless scalar, in agreement  
with (\ref {2lagr}). The limit $m_g\to 0$, however, 
is discontinuous, since for any nonzero $m_g$ in  (\ref {2lagr})
one has to satisfy  the Bianchi identity  which would  not appear as 
a constraint if  $m_g$ had been set to zero in (\ref {2lagr}) 
from the very beginning.

It is important to specify the limits of applicability of the above
condensation mechanism.
(I) The Lagrangian (\ref {lagr0}) could contain  a quartic 
interaction term  for the scalar $\lambda (\phi^* \phi)^2=\lambda\sigma^4/4$.
It is straightforward to check that our results will 
hold as long as $\lambda m_g^2 \ll g^2 m_H^2$.  
(II) The scalar could have an additional Yukawa  
term,  $q (\phi^*{\bar \psi}_1\Gamma {\psi}_2+ {\rm h.c.})$, 
where $q$ is a coupling, $\Gamma$ denotes either the ${\bf 1}$
or $i\gamma_5$ matrix depending on the spatial parity of $\phi$, and  
${\psi}_{1,2}$ denote fermions with different $U(1)$ charges that
render the Yukawa term gauge invariant.  One, or both of 
these fermions  could be setting the background charge density ${\bar J}_0$.  
The fermion  condensate,  
$\langle {\bar \psi}_1 {\psi}_2+ {\rm h.c.}\rangle$, if non-zero,  
could act as a source for the scalar. In order for this not to change 
significantly our results, the condition 
$q\langle {\bar \psi}_1 {\psi}_2+ {\rm h.c.}\rangle \ll m^2_H \sigma_c$
should be met\footnote{The  Yukawa coupling 
would also lead to the new terms in the fermion mass matrix. Depending 
on a concrete context, this may or may not impose 
additional constraints.}. (III) Due to the above Yukawa couplings
the scalar $\phi$ can decay. In order for the condensate phase to 
form in the first place, the ``condensation time'' $\sigma^{-1}_c$
has to be shorter then the lifetime of the $\phi$.
Thruough the work we will be checking the  
conditions (I-III) when appropriate. 

If the number density of the background fermions is such that it 
allows for the average inter-particle separation between them  
to be greater than the  Bohr radius of a fermion-scalar bound state, 
then, the fermions would  likely form a crystalline structure at 
low temperatures. If the resulting crystal is due to  the metallic bonding, 
that is it supports quantum gas of almost free scalars, 
then the condensation of the  scalars described above 
would be similar to the condensation of Cooper pairs in 
superconductors.  This case could be realized if $J_0 \lsim  g^6m_H^3$. 

On the other hand, if the average inter-particle separation between the 
background fermions is much smaller than  the would-be Bohr radius of 
the fermion-scalar bound state,  then the conventional 
quantum-mechanical considerations of the van der Waals, ionic, 
covalent or metallic bonding would not be applicable.
This would corresponds to the choice $J_0 \gsim g^6m_H^3$. 
In this case, the background fermions do not have to form an ordered 
structure, and yet, we'd expect the condensation of scalars. 
Moreover, the argument that the crystalline structure should be 
lost at some  high density  is supported by the discussions 
in a paragraph below.

A special sub-case of the discussion in the above paragraph 
is when $J_0 \gg m_H^3/g^6$:  It is straightforward to 
deduce  from the results obtained above  that the average inter-particle  
separation in the system, although is smaller than the would-be 
Bohr radius,  is greater than  the Compton wavelength of the  
massive photon.  If so, then, the electric charges
of the fermions and bosons are screened for all our purposes. 
The above described  condensation mechanism, with a good approximation, would 
reduce to the standard Bose-Einstein condensation of (almost) free scalars.
This system would behave as a two-component substance of free fermions 
and condensed scalars.

\vspace{0.1in}  

{\it 3. Finite-volume regularization.} Here we would like to regularize 
the infinite-volume theory of the previous section. Consider a material 
ball of a fixed radius $R$ which has a built in constant charge 
density $g{\bar J}_0$  uniformly distributed over its  volume. We will 
assume that such a ball  is prepared ``by hands'' with appropriate 
charges, and address the question: How does the electric 
potential  of this ball look like when the charged condensate 
described in the previous section compensates the fermion charge 
in its interior? This question is similar in spirit to the one 
we ordinarily  study for,  e.g., a uniformly charged insulating ball in 
electrodynamics.  

We'll be looking for static solutions of 
eqs. (\ref {Name}), which we parametrize as follows:
\beq
B_0(r) = B_{0c}+\delta B_0 (r)\,,~~~ \sigma(r) = \sigma_c +\delta \sigma(r)\,.
\label{exp}
\eeq
We focus on the solutions that in the interior of the ball 
satisfy $\delta \sigma/\sigma_c \ll 1$ and $\delta B_0/ B_{0c} \ll 1$. 
Then the equations for $\delta B_0 $ and $ \delta \sigma$ become:
\beq
-\nabla^2 \delta B_0 + m_g^2\, \delta B_0 = -2 \,m_g \,m_H \,\delta \sigma \,,
\label{beqn} \\
-\nabla^2 \delta \sigma = 2 \,m_g \,m_H \,\delta B_0 \,,
\label{seqn}
\eeq
where, as before,  $m_g \equiv g \sigma_c$.  
Explicit solutions of the above equations  can be readily found. 
For simplicity, we will present them for  $m_H \gg m_g$, i.e., when 
the $m_g^2\,\delta B_0$ term in the first equation can be neglected. 

The solutions in the interior of the ball are
\beq
\delta B_0(r) = \frac{1}{r} \left[c_1 \sinh(M r) 
\cos(M r)+c_2 \cosh(M r) \sin(M r)\right] \,,
\label{bsol} \\
\delta \sigma(r) = \frac{1}{r} \left[-c_1 \cosh(M r) 
\sin(M r) +c_2 \sinh(M r) \cos(M r)\right] \,,
\label{ssol}
\eeq
where $M \equiv \sqrt{m_g\,m_H}$, and $c_1$ and $c_2$ are 
constants to be determined from matching these solutions to the 
exterior ones.

Outside of the ball we approximate the solutions to be
\beq
B_0 = \frac{Q}{r}\,,~~~ \sigma = k \frac{e^{-m_H (r-R)}}{r} \, ,
\eeq
where $Q$ is a yet-unknown effective charge of the ball, which 
should be determined from the matching conditions, and which 
we expect to be mostly concentrated near the surface.  
By matching the solutions  and their first derivatives 
at $r = R$, we find
\beq
&c_1& = \frac{2}{gD} [m_g (m_H R +1) (\sinh(MR)\sin(MR)+\cosh(MR)\cos(MR)) \\
&+& m_H (\sinh(MR)\sin(MR)-\cosh(MR)\cos(MR)-{m_H\over M} 
\sinh(MR)\cos(MR))], 
\nonumber  \\
&c_2 &= \frac{2}{gD} [m_g (m_H R +1) (\sinh(MR)\sin(MR)-\cosh(MR)\cos(MR)) \\
&-& m_H (\sinh(MR)\sin(MR)+\cosh(MR)\cos(MR)+{m_H\over M} 
\cosh(MR)\sin(MR))]. 
\nonumber 
\end{eqnarray}
While, for the charge $Q$  we obtain the following 
expressions:
\beq
Q &=& \frac{1}{gD} [(m_g (m_H R +1)+m_H(m_H R-1))\sinh(2MR) \label {Q} \\
&-&(m_g (m_H R +1)-m_H(m_H R-1))\sin(2MR) \nonumber \\
&+&(2 m_H M R-m_H^2/M)\cosh(2MR)+(2 m_H M R+m_H^2/M)\cos(2MR)],
\nonumber
\eeq
where $ D \equiv m_H \sinh(2MR)+m_H \sin(2MR)+2M\cosh(2MR)+2M\cos(2MR).$
Finally, the constant $k$ is determined as
\begin{eqnarray}
k &=& \frac{1}{gD} [-(m_g+m_H)\sinh(2MR)-(m_g-m_H)\sin(2MR)\\
&+&2 m_g M R\cosh(2MR)+2 m_g M R\cos(2MR)] \nonumber \,.
\end{eqnarray}
In the case of physical interest, $MR \gg 1$,  the above  solutions 
have a number of interesting properties.  
The net charge density in the ball, $g{\bar J}_{0\,\rm eff} = g{\bar J}_0-g^2 
\sigma(r)^2 B_0(r)$, 
is exponentially small in the interior, except in a narrow spherical shell
near the surface of width $M^{-1}$.  Thus, the charge is screened in 
the bulk of the ball, but there remains an unscreened surface charge.  
In this limit the effective charge of the ball is 
$Q = m_H R/g = g{\bar J}_0 R^3 /(m_g R)^2$. This system is characterized
by the conserved electric charge $Q$, and conserved
fermion number $N= {\bar J}_0  R^3/3$.  

If we increase $R\to \infty$,  with all the other parameters held
fixed, the effective charge should also grow linearly with $R$
in order for the condensate phase to be possible inside the ball.
Put in other words, in order to prepare  a ball of a given 
radius with the charged condensate phase inside, one has to 
retain a specific amount of charge $Q$ defined in (\ref {Q}),
on its surface. Hence, in the infinite volume limit considered in the previous 
section,  there is ``a surface at infinity'' 
that carries charge.  This charge is responsible for 
the constant $B_0$ in (\ref {b0}).

In the bulk of the ball the electric field
and the electromagnetic energy are negligible. Closer to the boundary, 
however, the surface energy becomes non-zero due to the varying 
electric field. The resulting expression scales as
\beq
{\rm{Energy}}_E \propto {Q^2\over R}\propto {m_H^2 R\over g^2} \,.
\label{surfg}
\eeq
From our solutions it is also straightforward to get the scaling of  
the volume  energy well within the  ball; it reads as 
$\sim m_H {\bar J}_0 R^3$.

Let us consider  an example of a physical system in which the  
charged condensate could potentially be obtained. Suppose in a laboratory 
one could prepare a reservoir, or a trap, in which negatively 
charged electrons and   
positively charged helium-4 nuclei, with a net negative charge, 
could be put together. Consider densities of these particles high enough so 
that the average separation between 
the particles, $\sim {\bar J}^{-1/3}_0$, is smaller than the size 
of a helium atom, which we estimate for simplicity 
to be the Bohr radius  $\sim 1/(\alpha_{\rm em}m_e)$ 
($\alpha_{\rm em}$ denotes the fine-structure 
constant, and $m_e$ is the electron mass;  we still stay somewhat lower 
than nuclear densities).
As long as ${\bar J}^{1/3}_0 \gsim \alpha_{\rm em}m_e$ the helium atoms in the 
substance would not form. According to the discussion at the end of 
Section 2, at high-enough densities (but still somewhat 
below the nuclear ones)  we would not expect the crystalline 
structure to form either.  Can the charged condensate be formed in 
this system?  Strictly speaking, the calculations of the 
previous section are not directly applicable to this case,
because electrons are lighter than the helium-4 nuclei and 
averaging over the electron positions to calculate the photon mass 
may not be a good approximation.  In this case we would expect 
the photon mass squared to be determined by $g^2 {\bar J_0}/m_e$, 
instead of  $g^2 {\bar J_0}/m_H$, which should be 
applicable when the fermions are heavier than the scalars.   
We can introduce small temperature in the above system 
to see under what conditions the condensation would take place.
Once the thermal de Broglie wavelengths of the helium-4 nuclei have 
overlaps with each other, and as long at the photon Compton wavelength
is shorter than  the thermal de Broglie wavelength, the system can 
be treated as a macroscopic mode, or the condensate. 
The former condition, $\lambda_T \sim (1/m_HT)^{1/2}\gsim {\bar J}^{-1/3}_0$,
would suggest that $T\lsim 10^{-1}~{\rm eV}\sim  10^{-5}~{\rm K}$, 
while the latter, $1/m_g \lsim  \lambda_T$, would give a stronger  bound 
$T\lsim 10^{-5}~{\rm eV}\sim  10^{-9}~{\rm K}$ 
(we use $g^2 {\bar J_0}/m_e$ as the photon mass squared).
Temperatures reached in experiments on Bose-Einstein condensation of 
atoms are within this  range, see, e.g., \cite {BEC}. 

Let us look at other characteristics of this system in the condensate phase. 
Suppose the size of the sphere, 
or the trap we are dealing with, was $\sim 1m$. Then, the number 
of electrons and helium-4 nuclei would  have to be 
$N\gsim (\alpha_{\rm em}m_e)^3 (1m)^3\sim 10^{33}$ for helium atoms 
not to form. The total mass of these particles would be  
$\gsim 10^{6}~kg$. Moreover, the photon in this 
substance would acquire the mass $m_g \gsim 10^4~{\rm  eV}$, while 
the unbalanced charge of $\gsim  10^{16}$ units  would be 
residing  near the surface, in a narrow spherical shell  of 
size $\sim 1/\sqrt{m_H m_g}\sim  10~{\rm fm}$. (The electric field 
strength near the surface of such a sphere would be 
enough to ionize the air, so we assume that it's placed 
in a vacuum chamber).  

Propagation of light in the bulk of this substance would proceed with a 
delay caused by the induced photon mass $m_g$. 
For simplicity, we have considered  above
the system of a macroscopic size, but nothing prevents one
to look at much smaller systems, e.g., for a $1$ mm
size system the required number  of electrons and helium-4 nuclei 
would  have to be $N\sim 10^{22}$, and the mass of the system 
$\gsim 10^{-5}~kg$.

Suppose a ball of a fixed radius and charge 
determined by (\ref {Q}) with the charged condensate had been prepared. 
What happens if we gradually  bring to the ball's surface additional 
charges that would decrees or increase $Q$?  In terms of the theory
considered above, this would imply 
that we're adding a nonzero scalar chemical potential
term $\mu_s$,  as discussed in the comment (iii) on pages 3 and 4.
In this case, the value of  $\langle gB_0 \rangle $ inside 
the ball would change to maintain  the value of 
the effective chemical potential, $-\mu_s + \langle gB_0 \rangle$, 
to be equal to $m_H$. In this case, one should expect 
the relation (\ref {Q}) to be modified.

Before turning to the next section, let us comment on certain  
limiting cases.  If $m_H\to \infty$, for fixed and finite $R$, we would 
expect the scalar field to decouple and the 
solution to turn into the one for the potential of an insulating 
ball populated by a constant charge density, for which  
the potential equals to $g{\bar J}_0({R^2\over 2}-{r^2\over 6})$ inside,
and  to $g{\bar J}_0 ({R^3\over 3r})$ outside. On the other hand, this would 
imply that $\delta \sigma = -\sigma_c$. However, our expansion 
breaks down in this regime, and the solutions (\ref{bsol}) 
and (\ref{ssol}) are no longer applicable. In the full perturbative 
expansion, the l.h.s. of  equations (\ref{beqn}) and (\ref{seqn}) include
the non-linear terms
\beq
+g \,m_H\, \delta \sigma^2+2 \,g\, m_g \,\delta \sigma \,
\delta B_0 + g^2\, \delta \sigma^2 \,\delta B_0 \,,
\label{bnonlin} \\
-g \,m_g\, \delta B_0^2-2 \,g\, m_H \,\delta \sigma \,
\delta B_0 - g^2\, \delta \sigma \,\delta B_0^2 \,,
\label{snonlin}
\eeq
respectively.  When $\delta \sigma = -\sigma_c$ these terms become 
relevant, and in fact recover the standard electrodynamics result:
$-\nabla^2 B_0 = g {\bar J}_0$. Moreover, at some point 
when  $m_H$ exceeds the background fermion mass,  mobility of 
the fermions will play a role and, in general,
our results should not be immediately applicable.

Alternatively, we could look at the limit in which 
$m_g \rightarrow 0$ for a fixed $m_H$,  i.e.,  ${\bar J}_0 \rightarrow 0$.  
In this case we have a massless photon and a massive scalar, with
$\sigma$ scaling as $m_g$.  Since this implies that 
$\delta \sigma \rightarrow  -\sigma_c$, the same argument
as above applies and the solutions (\ref{bsol}) and (\ref{ssol}) 
are not applicable. 

Finally, in the limit $m_H \rightarrow 0$ we return back to 
equations (\ref{beqn}) and (\ref{seqn}) and now take $m_H \ll m_g$
so that we neglect the r.h.s. of the first equations.   Then, 
it would  seem that as $m_H \rightarrow 0$ the solutions approach the 
trivial ones,  $B_0=0$ and $\sigma = 0$. To see how we arrived at this 
erroneous result we again return to the non-linear terms (\ref{bnonlin}) 
and (\ref{snonlin}) which become significant in this limit.
Retaining these terms in our equation for $B_0$, we set $\sigma = 0$ 
and recover the expected electrodynamics result.

In the present work we left out a question of existence of 
a soliton with the charged condensate phase inside, 
that would be stable due to sufrace effects. Such an object would be
somewhat  similar to a droplet in a liquid drop model of 
the nucleus (see, e.g.,  \cite{Resnik}).  The related issues  will 
be discussed in \cite {GGRR}.

\vspace{0.1in}

{\it 4. Comments on compact objects.} In in this section we 
will use the power of gravity as a stabilizer to suggest  a
possible manifestation  of the charged condensation in astrophysics.  
We consider compact objects. In a general setup,  due to 
energy considerations, the condensing scalar would be a lightest 
charged  scalar available in 
the spectrum \cite{GGRR}, that could condense before decaying. 
If no new light charged scalars exist, then a first candidate 
would be a charged pion. However, in order for pions not to decay,   
one should consider high  densities, e.g., the conditions similar  
to the ones for pion condensation  in neutron stars \cite {Migdal}.
 
Charged condensate in compact objects with electrons and helium-4 nuclei 
could also exist. These object could be  
held together by  gravity which is competing 
against the degeneracy pressure of the fermions\footnote{This 
is similar to the stabilization mechanism in white dwarfs and  
neutron stars.}.  Since this mechanism is generic, 
and since we would expect any such object  
to contain a mixture of various species, we will discuss it 
in general terms of background fermions and charged scalars. 

Consider a distribution of $N$ charged fermions and $N_s$ 
charged scalars  with the net electric charge $Q$.  Such a 
distribution could collapse under the influence of gravity
into a compact object, a droplet. Below we consider a regime in which 
gravitational force is dominating over the electrostatic forces 
at the surface of the droplet. Moreover, we will assume that 
the temperature in the interior is low enough for all particles to be 
treated non-relativistically. Then, at a certain temperature, 
there should be a phase  transition in the interior into the 
charged condensate state. At that point the relation 
$\langle gB_0\rangle -\mu_s(T_c) = m_H(T_c)$ will be 
satisfied.  

To get qualitative estimates of the size of such a droplet we 
will ignore the difference between the values of $N$  and $N_s$, and 
minimize  energy as a function of the radius $R$  at a fixed 
value of the charged particle number $N$. Since these discussions 
are qualitative, we'll be omitting the factors of order 
10 or less. The total energy of a droplet reads:
\beq
{{E(R)}} =  m_H N + N\sqrt{p_J^2+m_J^2} - \frac{GM^2}{R},
\label{Evar}
\eeq 
where the first term is the energy of the condensate; 
the second term is the energy 
of a non-interacting  gas of charged particles that give rise
to the background density $J_0$ (hence, the subscripts 
in $p_J,m_J$); and the last term is
due to gravity, where $G$ denotes the Newton's constant 
(we'll be using the Planck mass 
$\mpl \equiv G^{-1/2}$), and $M$ is the total mas of the 
droplet  which depends on $N$. We have ignored in (\ref {Evar}) 
the surface terms which are negligible in the regime where gravity 
is dominant.

The critical radius reads: $R_c \sim {B^2/m_J N^{1/3}}$
where $B \equiv {\mpl /  (m_H +m_J)}$. This 
leads  to the  expression for the critical energy
\beq
{{E_c}} = (m_H +m_J) N \left[ 1 -\left(\frac{m_J}{m_H}\right) 
\left ( {N^{1/3} \over B} \right )^4 \right] \,.
\label{EcB}
\eeq
The critical radius decreases with increasing
$N$, the bounds on which are:
\beq
\frac{1}{e^{1/2}} \left(\frac{m_H}{m_J}\right)^{3/4} 
B^{9/4} \lsim N \lsim 
{\rm min} \left \{
\left ({m_H\over m_J}
\right )^{3/4} 
B^3; ~~  B^3\,
\right \}.
\label{NB}
\eeq
Here the lower bound is due to the requirement that gravity be dominant 
in stabilizing this object, and the upper bound is for the relativistic 
gravitational and fermionic effects to be negligible. These objects 
are stable as long as the gravitational binding energy in (\ref {EcB}) 
exceeds the electrostatic energy of uncompensated charges on its
surface. This constraint is taken into account by the 
bounds (\ref {NB}).

In a simple case when the droplet is assumed to be  made of electrons 
and the charged condensate of helium-4 nuclei, $N$ has to be  
close to the upper bound in (\ref {NB}), $N\sim 10^{57}$. 
The mass of this object is within an order of magnitude of 
the mass of the Sun,  and its size is $\sim 10^{6}~m$.
This object has characteristics that are  similar to 
those of neutron stars (except that it will have some surface 
charge, that was negligible in our considerations).  However, 
propagation  of light through such a cold and dense 
object will have specific characteristics described 
in Sections 2 and 3.

\vspace{0.2in}

{\it Acknowledgments.} We'd like to thank  R. Barbieri, 
V. Berezhiani, Z. Chacko,  G. Dvali, M. Kleban, I. Klebanov, 
M. Laine, S. Mukhanov and N. Weiner for useful discussions. The work 
of GG is supported by NASA grant NNGG05GH34G and NSF grant 0403005.
RAR is supported by  James Arthur graduate fellowship.

\vspace{0.2in}



\begin{thebibliography}{99}


\bibitem{BEC} M.~H.~Anderson, J.~R.~Ensher, M.~R.~Matthews, 
C.~E.~Wieman and E.~A.~Cornell,
  Science {\bf 269} (1995) 198; 
K.~B.~Davis, M.~O.~Mewes, M.~R.~Andrews, N.~J.~van Druten, 
D.~S.~Durfee, D.~M.~Kurn and W.~Ketterle,
  Phys.\ Rev.\ Lett.\  {\bf 75} (1995) 3969.

\bibitem{Migdal} A.B. Migdal, Zh. Eksp. Theor. Fiz. {\bf 61} (1971) 2210,
[Sov. Phys. JETP {\bf 34} (1972) 1184;  R.~F.~Sawyer and D.~J.~Scalapino,
  Phys.\ Rev.\  D {\bf 7}, 953 (1973).

\bibitem{Linde} A.~D.~Linde,
  Phys.\ Rev.\  D {\bf 14}, 3345 (1976).

\bibitem{Kapusta} J.~I.~Kapusta,
  Phys.\ Rev.\  D {\bf 24} (1981) 426.

\bibitem{GGRR} G. Gabadadze and  R. A. Rosen, in progress. 

\bibitem{Resnik} R. Eisberg, R. Resnick, ``Quantum Physics of Atoms, 
Molecules, Solids, Nuclei, and Particles'', John Wiley \& Sons, Inc. 1985.


\end{thebibliography}
\end{document}